\documentclass[12pt]{article}
\usepackage{amsmath,amssymb,slashed}
\usepackage{amssymb,amsmath,bm}
\usepackage{color}
\usepackage{graphicx}

\newcommand{\as}{\alpha_s}
\newcommand{\eps}{\epsilon}
\newcommand{\mupi}{\mu_\pi^2}
\newcommand{\mug}{\mu_G^2}
\newcommand{\rd}{\rho_D^3}
\newcommand{\rls}{\rho_{LS}^3}

\newcommand{\ep}{\epsilon }
\newcommand{\uh}{{\hat u} }

\newcommand{\MS}{\ensuremath{\overline{\text{MS}}}}

\long\def\symbolfootnote[#1]#2{\begingroup%
\def\thefootnote{\fnsymbol{footnote}}\footnote[#1]{#2}\endgroup}

\newcommand\gsim{\mathop{\mbox{\vbox{\hbox{$>$} \vskip -9pt \hbox{$\sim$}
             \vskip -3pt  }}}}

\def \be{\begin{equation}}
\def \ee{\end{equation}}
\newcommand{\bea}{\begin{eqnarray}}
\newcommand{\eea}{\end{eqnarray}}
\def \nn{\nonumber}

\numberwithin{equation}{section}
\allowdisplaybreaks[4]

\begin{document}
\begin{titlepage}

\vspace*{3.5cm}

\begin{center}{\Large\bf\boldmath Perturbative corrections to power suppressed\\[3mm] effects in 
semileptonic $B$ decays}
\end{center}
\vskip 2.5cm

\begin{center}
  {\bf Andrea Alberti$^{a}$, Paolo Gambino$^a$, %\\[2mm]
   Soumitra
    Nandi$^{b}$} \\[5mm]
  {\sl $^a$\,Universit\`a di Torino, Dip.\ di  Fisica \& INFN Torino, I-10125, Italy}\\[2mm]
        {\sl $^d$\,Dept.\ of Physics,
Indian Inst.\ of Technology Guwahati, 781 039, India}
\end{center}

\vskip 3cm

\begin{abstract}
  We compute the $O(\alpha_s)$ corrections to the Wilson coefficient of the chromomagnetic operator 
  in inclusive semileptonic $B$ decays. The results are employed to evaluate the complete $\as\Lambda_{QCD}^2/m_b^2$ correction to the semileptonic width and to the first moments of the lepton energy distribution.

  \end{abstract}

\end{titlepage}

\section{Introduction}
The model-independent study of inclusive semileptonic $B$ decays, initiated twenty years ago \cite{Bigi:1992su,Blok:1993va}, is based on an Operator Product Expansion (OPE) in conjunction with the heavy quark expansion.
At the B factories, it 
has allowed for very precise determinations of the CKM matrix elements $|V_{cb}|$ and  $|V_{ub}|$, mostly limited by theoretical uncertainties \cite{hfag}.
Further progress therefore requires theoretical improvements, as well
as high statistics data from Belle-II in the case of charmless decays.   The calculation of higher order corrections in the OPE, in particular, is  of crucial importance.

 The OPE  expresses  the  widths and the first moments of the kinematic distributions of $B\to 
 X_{u,c} \ell\nu$ as double expansions in
$\as$ and $\Lambda_{\rm QCD}/m_b$.  
 The leading terms in these double expansions are 
given by the free $b$ quark decays, while  the  $O(\alpha_s,\alpha_s^2 \beta_0)$  perturbative corrections \cite{pert,Aquila}  and the $O
(\Lambda^2_{\rm QCD}/m_b^2, \Lambda^3_{\rm QCD}/m_b^3)$ non-perturbative corrections \cite{Blok:1993va,1mb3} have been known for a long time. More recently, the complete $O(\alpha_s^2)$
calculation has been completed \cite{alphas2}, and the $O((\Lambda_{\rm QCD}/m_Q)^{4,5})
$ have been investigated \cite{Mannel:2010wj}. 
The parameters of the double expansions are the heavy quark masses $m_b$ 
and  $m_c$, the strong 
coupling $\alpha_s$, and the $B$-meson matrix elements of  local operators of growing 
dimension. The latter parameterize all the long-distance physics that is relevant for  inclusive decays:
at $O(\Lambda_{\rm QCD}^2/m_b^2)$ there are two parameters,  $\mupi$ and $\mug$, at
  $O(\Lambda_{\rm QCD}^3/m_b^3)$ two more appear, 
 $\rd$  and  $\rls$, and so on.  The non-perturbative parameters  are constrained by 
the experimental data for the moments of the lepton energy and hadron mass distributions of $B\to X_c \ell \nu$ and can be employed to extract $|V_{cb}|$ from the semileptonic width. Recent fits can be found in Refs.~\cite{hfag, newfit}.

The coefficients of the non-perturbative corrections of  $O(\Lambda^n_{\rm QCD}/m_b^n)$ in the double series are Wilson coefficients of 
power-suppressed local operators and can be computed perturbatively.
Only a subset of the $O(\alpha_s\Lambda_{\rm QCD}^2/m_b^2)$ corrections has been computed so far:
the $O(\as)$ corrections to  the coefficient of  $\mupi$ \cite{Becher:2007tk, 
Alberti:2012dn}, which represents   the 
$B$ meson expectation value of the kinetic operator and is related  to the average kinetic energy of  the $b$ quark in the $B$ meson. In this paper we present the  calculation of the remaining 
$O(\alpha_s\Lambda_{\rm QCD}^2/m_b^2)$  corrections, those proportional to $\mu^2_G$,
the expectation value of the chromomagnetic operator. We compute the corrections to the triple differential semileptonic $B$ decay width and therefore to the most general moment, in such a way that they can be readily employed 
to improve the precision of the  fits to $|V_{cb}|$.

Our calculation follows the method outlined in Ref.~\cite{ewerth}, where the same corrections
were computed in the simpler case of $B\to X_s \gamma$, and  in Ref.~\cite{Alberti:2012dn}. Here we discuss  the matching procedure in greater detail and present analytic results for the $O(\alpha_s\mu_G^2/m_b^2)$ corrections to the three relevant structure functions. In this way the corrections to the triple differential width become available  and the corrections to arbitrary moments can be  computed. We then present numerical results for the semileptonic width and for the first leptonic moments.
The paper is organized as follows: after setting the notation in Section 2, 
we discuss the matching in Section 3; the following Section presents and discusses the 
numerical results. Section 5  summarizes our findings.
The lengthy analytic results for the structure functions are given in the Appendix.

\section{Notation}
We consider the decay of a $B$ meson 
of four-momentum $p_B= M_B v$ into a lepton pair with momentum $q$ and 
a hadronic final state containing  a charm quark with momentum $p'=p_B-q$.
The hadronic tensor $W^{\mu\nu}$ which determines the hadronic contribution to the differential 
width  is given by the absorptive part of a current correlator in the appropriate kinematic region,
\be
W^{\mu\nu}(p_B,q)= {\rm Im} \,  \frac{2\,i}{\pi M_B}  \int d^4 x\, e^{-i q\cdot x} \langle \bar B| T J_L^{\mu\dagger}(x) J_L^{\nu} (0) | \bar B\rangle,
\label{correlator}
\ee
where $J_L^\mu=\bar c \gamma^\mu P_L b$ is the charged weak current. The correlator admits an OPE in terms of local operators, which at the level of the 
differential rate takes the form of an 
expansion in inverse powers  of the energy  release, whose leading term corresponds to the 
decay of a free quark. 

Our notation follows that of Ref.~\cite{Aquila,Alberti:2012dn}. We  express the $b$-quark decay kinematics in terms of the dimensionless quantities 
\be
\rho= \frac{m_c^2}{m_b^2}, \qquad\quad \hat u= \frac{(p-q)^2 - m_c^2}{m_b^2} ,\quad\qquad \hat q^2= \frac{q^2}{m_b^2},
\ee
where $p= m_b v$ is the momentum of the $b$ quark and 
\be
0\le \hat u \le \hat u_+ =(1-  \sqrt{\hat{q}^2})^2 -\rho \qquad  {\rm and} \qquad 
0\le \hat q^2 \le (1-\sqrt{\rho})^2.
\label{physrange}
\ee
The energy of the hadronic system, normalized to the $b$ mass, is 
\be
E=  \frac12 (1+\rho +\hat u -\hat q^2).
\ee
Tree-level kinematics correspond to $\hat u=0$, in which case we indicate the energy of the hadronic final state as $E_0$.
 The  normalized total leptonic energy is $\hat q_0 =1-E$ from which  follows 
$\hat u = 2\,(1-E_0-\hat q_0)$.
It is customary to decompose the hadronic tensor
as follows
\be
m_b \,W^{\mu\nu}(p_B,q)=-W_1 \, g^{\mu\nu}+W_2 \,v^\mu v^\nu +i W_3 \, \ep^{\mu\nu\rho\sigma }
v_\rho \hat q_\sigma + W_4 \hat q^\mu \hat q^\nu +W_5 \left(v^\mu \hat q^\nu\!+\! v^\nu \hat q^\mu\right),\label{eq::dec}
\ee
where  the structure functions $W_i$ are functions of $\hat q^2, \hat q_0$ or equivalently of
$\hat q^2, \hat u$, $v^\mu$ is the four-velocity of the $B$ meson, and $\hat q^\mu=q^\mu/m_b$.  As  only $W_{1,2,3}$ contribute to the decay rate for massless leptons, we will concentrate on these three structure functions. 

Due to the OPE, the structure functions can be expanded in series of $\alpha_s$ and $\Lambda_{\rm QCD}/m_b$. There is no term linear in $\Lambda_{\rm QCD}/m_b$ and therefore
\be\label{Wi}
W_i =W_i^{(0)} + \frac{\mu_\pi^2}{2m_b^2} W_i^{(\pi,0)}+\frac{\mu_G^2}{2m_b^2} W_i^{(G,0)}+
  \frac{\as}{\pi}\left[ C_F W_i^{(1)} + C_F
  \frac{\mu_\pi^2}{2m_b^2} W_i^{(\pi,1)}+\frac{\mu_G^2}{2m_b^2} W_i^{(G,1)}\right]
\ee
where we have neglected terms of higher order in the expansion parameters.
$\mu_\pi^2$ and $\mu_G^2$ are the $B$-meson matrix elements of the only gauge-invariant 
dimension 5 operators that can be formed from the $b$ quark and gluon fields:
\begin{equation}\label{eq::lambdai}
  \mu_\pi^2 = \frac{1}{2M_B}\langle\bar B|\bar b_v(i\vec D)^2b_v|\bar B\rangle\,,\qquad
  \mu_G^2 = -\frac{1}{2M_B}\langle\bar B|\bar b_v\frac{g_s}{2}G_{\mu\nu}\sigma^{\mu\nu}b_v|\bar B\rangle\,,
\end{equation}
where $b_v$ is the static quark field, and $G_{\mu\nu}= G_{\mu\nu}^a T^a$ is the gluon field tensor, which is defined as 
$g_sG^a_{\mu\nu}T^a=-i[D_\mu,D_\nu]$ with the covariant derivative  $D_\mu=\partial_\mu+ig_sG_\mu^aT^a$.\footnote{Since we are only interested in $\Lambda_{\rm QCD}^2/m_b^2$ corrections, $\mu_\pi^2$ and $\mu_G^2$ are here defined in the asymptotic HQET regime, {\it i.e.} in the infinite mass limit.}
The leading order coefficients are given by 
\be
W_i^{(0)} = w_i^{(0)} \,\delta(\hat u);  \qquad  \qquad w_1^{(0)} = 2 E_0, \qquad w_2^{(0)} = 4, \qquad w_3^{(0)} =2.
\ee
The tree-level and one-loop coefficients of $\mu_\pi^2$ can be found in \cite{Alberti:2012dn};
the tree-level coefficients  of $\mu_G^2$  \cite{Blok:1993va}, using  $\lambda_0=4(E_0^2-\rho)$,  can be written as: 
\be
W_i^{(G,0)} = w_i^{(G,0)} \,\delta(\hat u) +w_i^{(G,1)} \,\delta'(\hat u);  \label{WiG}
\ee
 \begin{center}
 \begin{tabular}{ll}
$w_1^{(G,0)}= -\frac43 (2-5E_0),$ \quad& $ w_1^{(G,1)}= -\frac43 (E_0+3E_0^2+\frac12 \lambda_0);$ \nn\\[1mm]
$w_2^{(G,0)}= 0, $\quad&$ w_2^{(G,1)}=\frac83 (3-5E_0)   ; $ \nn\\[1mm]
$ w_3^{(G,0)}=\frac{10}3 ,$ \quad&$ w_3^{(G,1)}= -\frac43 (1+5E_0) .$\nn
\end{tabular}
\end{center}

\section{The matching at $O(\as)$}
Schematically, we can write  the OPE in momentum space as
\be\label{eq::ope}
\frac{2i}{\pi}\int d^4 x \,e^{-i q\cdot x}\,T[ J_L^{\dagger\mu}(x) J_L^\nu(0)] = \sum_i 
%\frac1{m_b^{d_i-3}}\, 
c^{(i)\mu\nu}_{\ \ \{ \alpha\} }(v,q) \,O_i^{\{\alpha\}}(0),
\ee
where %$d_i$ is the canonical dimension of 
$O_i^{\{\alpha\}}$ are  local operators  and $\{\alpha\}$ stands for possible additional Lorentz indices. The number of local operators of dimension $d_i\le 5$ that contribute to the rhs
can be reduced, and their renormalization simplified, by resorting to the Heavy Quark 
Effective Theory (HQET) and using the relation between  the HQET static quark $b_v$ and 
the  QCD $b$ field,
\be
b(x)= e^{-i m_b v\cdot x}\left(1+\frac{i \slashed{D}}{2m_b} \right) b_v(x) \, .
\ee
Eventually,  we will  need the following  set:
\begin{align}\label{eq::operators}
  &\qquad\qquad\qquad\qquad O_b^\mu = \bar b\gamma^\mu b, \hspace{3cm} O_s= \bar b \,b\, , &\nonumber\\[1mm]
  &O_1^{\mu} = \bar b_viD^\mu b_v, \qquad 
  O_2^{\mu\nu} = \bar b_v\frac{1}{2}\{iD^\mu,iD^\nu\}b_v\,, \qquad
  O_3^{\mu\nu} = \bar b_v\frac{g_s}{2}G^{\mu}_{\ \alpha}\sigma^{\alpha\nu} b_v\,.&
\end{align}
%where %$D^\mu$ is the covariant derivative and $g_sG^a_{\mu\nu}T^a=-i[D_\mu,D_\nu]$. 
Notice that $O_{b,s}$ are written in terms of the
QCD bottom quark field, while the other operators are constructed in terms of $b_v$.
Up to terms of dimension six, the operator $O_s$ can be expressed in terms of the 
others:
\be 
O_s= v_\mu O^\mu_b + \frac{O_{2\,\alpha}^\alpha+ O_{3\,\alpha}^\alpha}{2m_b^2}  + O\left(\frac1{m_b^3}\right)
\ee
but we keep it distinct for reasons that will become clear.
 We also find  operators that include a $\gamma_5$, but they can be neglected in our discussion. Indeed,  
because of the parity invariance  of strong interactions,
only the operators  in (\ref{eq::operators}) have non-vanishing matrix elements in the $B$ meson.  
 As we perform an off-shell calculation, we have not  used the HQET equation of motion for the $b_v$ field, which would reduce  the operator $O_1^\mu$  to a linear combination of $O_{2,3}^{\mu\nu}$. The equation of motion will be used only in the last step of the calculation, when we evaluate the matrix elements of the operators in the  $B$ meson.

In order to determine the Wilson coefficients $c^{(i)\mu\nu}_{\ \{\alpha\}}$ we compute renormalized Green's functions of both sides
of Eq.~(\ref{eq::ope}) on  heavy quark states close to the mass shell. The external heavy quarks have  residual momentum $k$ and we Taylor expand the Green's functions for small $k$ up to  to second order. 
To extract $c^{(3)\mu\nu}_{\ \ \alpha\beta}$  we  also need to consider  Green's functions with 
a soft  external gluon. They are  Taylor expanded  in both $k$ and the gluon virtuality $r$.

\begin{figure}[t]
\begin{center}
\includegraphics[width=13cm]{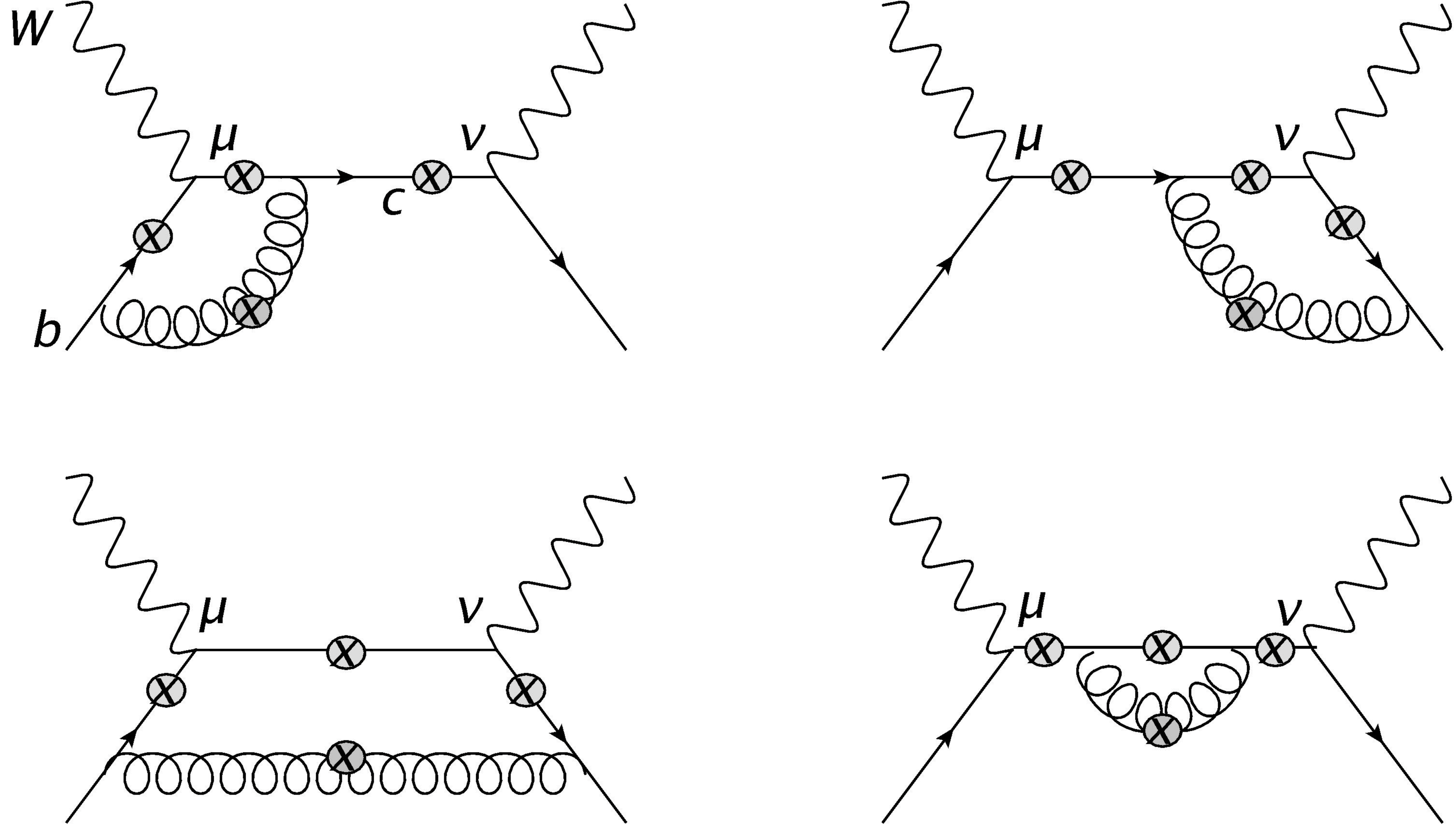}
\caption{\sf One-loop diagrams contributing to the current correlator. The background gluon can be attached wherever a  cross is marked.}
\label{fig1}
\end{center}
\end{figure}

It is convenient to decompose the tensors as  in (\ref{eq::dec}), writing the lhs  of Eq.~(\ref{eq::ope})as 
\be\label{tensor}
 T_{\mu \nu}= \frac1{m_b}\left[-g_{\mu \nu} T^{(1)} +v_{\mu}v_{\nu} T^{(2)} -i \epsilon_{\mu \nu \alpha \beta} v^{\alpha} \hat q^{\beta} T^{(3)} +\hat q_{\mu}\hat q_{\nu} T^{(4)} +(v_{\mu}\hat q_{\nu}+\hat q_{\mu}v_{\nu}) T^{(5)} \right].
\ee 
%The decomposition does not interfere with our subsequent considerations. 
For massless leptons, only 
the first three form-factors, $T^{(1-3)}$, contribute to  physical quantities.  
 Eq.~(\ref{eq::ope}) becomes
\be\label{eq::ope2}
T^{(i)} = c^{(i,b)}_{\ \  \alpha}  \,O_b^{\alpha} +
c^{(i,s)}  \,O_s+   %\frac1{m_b} \,
c^{(i,1)}_{\ \ \alpha}  \,O_1^{\alpha} +   %\frac1{m_b^2} \,
c^{(i,2)}_{\  \alpha \beta} \,O_2^{\alpha \beta} +  %\frac1{m_b^2} \,
c^{(i,3)}_{\ \alpha \beta} \,O_3^{\alpha \beta}+ \dots,
\ee
where the ellipses stand for contributions of operators of canonical dimension six or higher.
All the Wilson coefficients can be expanded in powers of  $\as$,
\be 
c_{\{\alpha\}}^{(i,m)}= c_{\{\alpha\}}^{(i,m,0)} + \frac{\as}{4\pi} \,c_{\{\alpha\}}^{(i,m,1)}+O(\as^2)\nonumber
\ee
and we are only interested in their imaginary part, cfr (\ref{correlator}). 
We consider the forward matrix element of (\ref{eq::ope2}) between two $b$ quarks, and between two quarks and a soft gluon:
\be\label{eq::bb}
\langle  T_i \rangle_{b\bar b}= c^{(i,b)}_{\ \  \alpha}  \,\langle O_b^{\alpha}\rangle_{b\bar b} + 
c^{(i,s)}  \,\langle O_s\rangle_{b\bar b} +
c^{(i,1)}_{\ \ \alpha}  \,\langle O_1^{\alpha}\rangle_{b\bar b} +   %\frac1{m_b^2} \,
c^{(i,2)}_{\  \alpha \beta} \,\langle O_2^{\alpha \beta} \rangle_{b\bar b}+  %\frac1{m_b^2} \,
c^{(i,3)}_{\ \alpha \beta} \,\langle O_3^{\alpha \beta}\rangle_{b\bar b}+ 
\dots,
\ee
\be\label{eq::bbg}
\langle  T_i \rangle_{b\bar bg}= c^{(i,b)}_{\ \  \alpha}  \,\langle O_b^{\alpha}\rangle_{b\bar bg} +  
c^{(i,s)}  \,\langle O_s\rangle_{b\bar bg} +
c^{(i,1)}_{\ \ \alpha}  \,\langle O_1^{\alpha}\rangle_{b\bar bg} +   %\frac1{m_b^2} \,
c^{(i,2)}_{\  \alpha \beta} \,\langle O_2^{\alpha \beta} \rangle_{b\bar bg}+  %\frac1{m_b^2} \,
c^{(i,3)}_{\ \alpha \beta} \,\langle O_3^{\alpha \beta}\rangle_{b\bar bg}+ 
\dots.
\ee
Here all the matrix elements should be interpreted as renormalized amputated Green's functions, either in full QCD (the lhs and the matrix elements of $O^\mu_b$ and $O_s$)
or in HQET;   since the two theories have the same infrared behavior the cancellation of infrared divergences is guaranteed.  The matrix elements  of a generic operator $O_X$ can be expanded in powers of $\as$,
\be
\langle O_X\rangle_{b\bar b(g)}=\langle O_X\rangle_{b\bar b(g)}^{(0)}+\frac{\as}{4\pi} 
\langle O_X\rangle_{b\bar b(g)}^{(1)}+ O(\as^2).\nonumber
\ee
We observe that 
\be
\langle O_s\rangle_{b\bar b}^{(0)}=\langle O_{3}^{\alpha\beta} \rangle_{b\bar b}^{(0)}=0\,  , \qquad\,
\langle O_b^{\alpha}\rangle_{b\bar bg}^{(0)}=\langle O_s\rangle_{b\bar bg}^{(0)}=0. \label{tl}
\ee
Therefore, at the tree-level, the expansion in the residual momentum $k$ of 
the lhs  of (\ref{eq::bb}) allows for the determination of $c^{(i,b,0)}_{\ \mu}$  at $k=0$,
of  $c^{(i,1,0)}_{\ \mu}$  at $O(k)$, of  $c^{(i,2,0)}_{\ \mu\nu}$  at $O(k^2)$.
More precisely, the $O(k)$ term in the lhs of   (\ref{eq::bb}) 
%and the $O(k^0)$ contribution to  (\ref{eq::bbg}) are 
is related  to the matrix elements of 
\be
\bar b \,\gamma^\alpha (i D^\beta - m_b v^\beta) b = 
 v^\alpha \,O_1^\beta + \frac1{m_b}\big( O_2^{\alpha\beta}+ O_3^{\alpha\beta}\big)
+O\Big(\frac1{m_b^2}\Big). \label{eq:m1}
\ee 
The latter equality  follows from the relation between $b$ and $b_v$ fields, and therefore the $O(k)$ term in the lhs of   (\ref{eq::bb}) 
contributes to the Wilson coefficients of $O_{1,2,3}$.
%The lhs of (\ref{eq:m1}) contribute also to the matrix elements with an external gluon.

For what concerns the Taylor expansion in $k,r$ of the lhs of (\ref{eq::bbg}),
the term at $k=r=0$ allows 
 for the determination of $c^{(i,1,0)}_{\ \mu}$, while the term linear in $k$ and $r$
determines    $c^{(i,2,0)}_{\ \mu\nu}$  and   $c^{(i,3,0)}_{\ \mu\nu}$. 
Gauge invariance guarantees that the same $c^{(i,1,0)}_{\ \mu}$ and $c^{(i,2,0)}_{\ \mu\nu}$
are extracted from the diagrams with and without external gluon. From (\ref{tl}) we also have 
$c^{(i,s,0)}=0$.

We write down explicitly the  tree-level coefficients only in the case of  $W_1$, 
namely for the first of the tensor structures in (\ref{tensor}) --- the other form factors have the same structure. We work in $d=4-2\epsilon$ dimensions and retain $O(\eps)$ terms 
\bea\label{eq::coeff0}
 {\rm Im\,} c^{(1,b,0)}_{\ \mu} &=& (1-\epsilon)\,  (v_\mu-\hat q_\mu) \, \delta (\hat u ) \\
 {\rm Im\,}c^{(1,1,0)}_{ \ \mu} &=&  \frac1{m_b} (1-\epsilon) \big[ 2\, (1-\hat q_0)\, 
 (v_\mu-\hat q_\mu) \,
\delta' (\hat{u} )+  v_{\mu}  \delta (\hat{u} ) \big]\\
{\rm Im\,}c^{(1,2,0)}_{\ \mu\nu} &=&  \frac2{m_b^2}(1-\epsilon) \, (1-\hat q_0) \,
\hat p'_\mu \,\hat p'_\nu
\,\delta''(\hat u)\nonumber\\
&&+ \frac2{m_b^2}(1-\epsilon)\left[\frac{1-\hat q_0}2 g_{\mu\nu}+2 v_\mu v_\nu-
 \frac32(\hat q_\mu v_\nu+ v_\mu \hat q_\nu)+ \hat q_\mu \hat q_\nu \right]\delta'(\hat u)\nonumber\\
&&-\frac1{m_b^2}\left[\epsilon g_{\mu\nu}-\frac{\hat q^2 v_\mu v_\nu -\hat q_0(\hat q_\mu v_\nu+ v_\mu \hat q_\nu) +\hat q_\mu \hat q_\nu}{\hat q^2-\hat q_0^2}\right]\delta(\hat u)
\\
{\rm Im\,}c^{(1,3,0)}_{\ \mu\nu} &=& -\frac2{m_b^2}\left[\frac{1-\hat q_0}2 g_{\mu\nu}(1+\epsilon) + ((1-\epsilon) \hat q_\mu -2v_\mu) \hat p'_\nu
 +\frac{\hat q\cdot \hat p' \,v_\mu \hat q_\nu -v\cdot \hat p' \,\hat q_\mu \hat q_\nu}{\hat q^2-\hat q_0^2}
  \right]\delta'(\hat u) \nonumber
\\&&
-\frac1{m_b^2}\left[\epsilon g_{\mu\nu}-\frac{\hat q^2 v_\mu v_\nu -\hat q_0(\hat q_\mu v_\nu+ v_\mu \hat q_\nu) +\hat q_\mu \hat q_\nu}{\hat q^2-\hat q_0^2}\right] \delta(\hat u)
\eea
where $\hat p'=  v -\hat q$. The $O(\epsilon)$ terms depend on whether the tensor decomposition of $T^{\mu\nu}$ is performed in  four (as in our case) or $d$ dimensions.

Eventually, of course, we need to evaluate Eq.~(\ref{eq::ope}) in the $B$ meson: the corresponding matrix elements of the operators (\ref{eq::operators})  are given by 
\bea &&\label{eq::matrixel}
\frac1{M_B} \langle \bar B|O_b^\mu| \bar B \rangle=2 \, v^\mu, \nonumber\\
&&\frac1{M_B} \langle \bar B|O_s| \bar B \rangle=2 -\frac{\mu_\pi^2-\mu^2_G}{m_b^2} , \;\, \nonumber\\
&&
\frac1{M_B} \langle \bar B|O_1^\mu |\bar B\rangle=  \frac{\mu_\pi^2-\eta\,\mu^2_G(\mu)}{m_b}\, v^\mu, \\
&&\frac1{M_B}\langle \bar B|O_2^{\mu\nu} |\bar B\rangle=-\frac{2\, \mu^2_\pi}{d-1} \left(g^{\mu\nu}- v^\mu v^\nu\right),\nonumber \\
&&
\frac1{M_B}\langle \bar B|O_3^{\mu\nu} |\bar B\rangle=\frac{ 2 \,\mu^2_G}{d-1} \left(g^{\mu\nu}- v^\mu v^\nu\right), \nonumber
\eea
where we have neglected higher order power corrections and  introduced the factor 
\be
\eta = 1+2\left[C_F +\left(1+\ln \frac{\mu}{m_b}\right) C_A \right] \frac{\as}{4\pi}  
 \ee
 in order to take into account the \(O(\alpha_s )\) corrections to the HQET equation of motion, in the same manner as it has been done in \cite{ewerth}.
In the standard tree-level calculation \cite{Blok:1993va}, one computes directly the 
coefficients of $\mu_\pi^2$ and $\mu_G^2$. However, in order to perform the renormalization properly it is essential  to distinguish between the various operators whose matrix elements
contain $\mu_G^2$.
%$|\bar B(p)\rangle$ and $|\bar B(v)\rangle$ are the physical and HQET hadronic states, that differ by normalization and higher power corrections, 
%$|\bar B(p)\rangle=\sqrt{M_B}[|\bar B(v)\rangle+ O(1/m_b)]$. 
The evaluation of Eq.~(\ref{eq::ope}) in the $B$ meson leads, through Eqs.~(\ref{eq::coeff0}-\ref{eq::matrixel}), to the well-known $O(\Lambda_{\rm QCD}^2/m_b^2)$ corrections \cite{Blok:1993va}, see also Eq.~(\ref{WiG}).

The one-loop calculation of the current correlator requires the imaginary part of the diagrams shown in Fig.~1. We use dimensional regularization for both ultraviolet and 
infrared divergences and proceed exactly as described in Ref.~\cite{Alberti:2012dn}. 
The result of the Taylor expansion in $k$ and $r$
is reduced to the master integrals listed in the of  Appendix of the same paper.
We perform the calculation in an arbitrary $R_\xi$ gauge and use the background field  gauge for the external gluon.
The ultraviolet divergences of the diagrams in Fig.~1 are removed by standard on-shell
quark mass and wave function QCD renormalization, see \cite{ewerth}.
Notice that the
$b\bar b$ one-loop amplitude at $k=0$ contains terms that lead
to $c^{(i,s,1)}\neq 0$;  in other words, $O_s$ emerges naturally from the OPE
{\it before} one uses the heavy quark expansion, and its presence  is essential 
to verify that $c^{(i,1,1)}_{\ \mu}$ and $c^{(i,2,1)}_{\ \mu\nu}$ extracted from the diagrams with and without external gluon are the same, as dictated 
by gauge invariance. 

The rhs of  (\ref{eq::ope2}) receives $O(\as)$ contributions from both 
one-loop matrix elements of the effective operators and the one-loop Wilson coefficients.
However, the unrenormalized one-loop matrix elements of $O_{1-3}$ vanish in dimensional regularization because they reduce to massless one-loop tadpole diagrams .  
The case of  $O_b^\mu$ is different and will be explained in a moment.
Besides  the on-shell wave function renormalization of the $b$ and $b_v$
fields, we need the operator renormalization, which is performed in
the $\MS$ scheme, see \cite{ewerth}. In particular
\begin{align}
  \left[c_{b\mu}O_b^\mu\right]^{\rm bare} &= Z_b^{\rm OS}\,c_{b\mu}O_b^\mu\,, &
  \left[c_{2\mu\nu}O_2^{\mu\nu}\right]^{\rm bare} &=
    Z_{b_v}^{\rm OS}Z_{\rm kin}^{\MS,\mu\nu\alpha\beta}\,c_{2\mu\nu}O_{2\alpha\beta}\,, &&\nonumber\\[1mm]
  \left[c_{1\mu}O_1^\mu\right]^{\rm bare} &= Z_{b_v}^{\rm OS}  \,c_{1\mu}O_1^\mu\,, &
  \left[c_{3\mu\nu}O_3^{\mu\nu}\right]^{\rm bare} &=
    Z_{b_v}^{\rm OS}Z_{\rm chromo}^{\MS,\mu\nu\alpha\beta}\,c_{3\mu\nu}O_{3\alpha\beta}\,.
\end{align}
where 
\begin{align}
  Z_{\rm kin}^{\MS,\mu\nu\alpha\beta} &= g^{\alpha \mu} g^{\beta \nu}-C_F \frac{3-\xi}{\epsilon}
  \left(g^{\mu\nu}-2v^\mu v^\nu\right)v^\alpha v^\beta\,\frac{\as}{4\pi}+\dots\,\nonumber\\[1mm]
  Z_{\rm chromo}^{\MS,\mu\nu\alpha\beta} &= g^{\alpha \mu} g^{\beta \nu}+ \frac{C_A}{\epsilon}
  \left(g^{\mu\alpha}-v^\mu v^\alpha\right)g^{\nu\beta}\,\frac{\as}{4\pi}+\dots\,.
\end{align}
The Feynman gauge is obtained by setting $\xi=1$. 
It is easy to see that the renormalization of  $O_2^{\mu\nu}$ is irrelevant 
because the matrix element of $Z_{\rm kin}^{\mu\nu\alpha\beta} O_{2\,\alpha\beta}$ vanishes
at the order of the calculation. On the other hand, the $B$ matrix element of  $Z_{\rm chromo}^{\mu\nu\alpha\beta} O_{3\,\alpha\beta}$ is proportional to that of $O_{3}^{\mu\nu}$, which simplifies the calculation.
 The operator $O_s$ does not need
renormalization because it enters at the loop level only.
The one-loop matrix elements of  $O_b^\mu$  do not vanish: they have to be 
Taylor expanded in $k$ and $r$ and included in the calculation. 

Putting together all pieces we have verified that all infrared and ultraviolet divergences are canceled in the Wilson coefficients and that the latter are independent of the amplitude
from which they are extracted. We have also verified that the results, which we 
express in terms of coefficients of $\as \mu_{\pi,G}^2$ in Eq.~(\ref{Wi}), do not depend on the quantum gauge parameter $\xi$. The coefficients of $\mu_\pi^2$, $W_i^{(\pi,1)}$, agree with Ref.~\cite{Alberti:2012dn}.
The complete analytic results for $W_i^{(G,1)}$ are given in the Appendix.

\section{Numerical results}
 In this section we present a preliminary investigation of  the numerical relevance of the $O(\as \Lambda_{\rm QCD}^2/m_b^2)$ corrections, using for  the heavy quark masses the reference values
$m_b=4.6$ GeV and $m_c=1.15$ GeV. First, we  consider on-shell quark masses; in this case the phase space integration of the triple differential width (see {\it e.g.}\ Eq.\,2.10 of Ref.~\cite{Alberti:2012dn})
leads to the total semileptonic width
\be
\Gamma_{B\to X_c \ell \nu}=\Gamma_0\left[ \left(1-1.78\, \frac{\as}{\pi}\right)\left(1-\frac{\mu_\pi^2}{2m_b^2}\right) - \left(1.94 +2.42  \,\frac{\as}{\pi}\right)\frac{\mu_G^2(m_b)}{m_b^2} \right],\nonumber
\ee
where $\Gamma_0= G_F^2 m_b^5 (1-8\rho+8\rho^3 -\rho^4 -12\rho^2 \ln\rho)/192 \pi^3$ is the tree level width, $\rho=m_c^2/m_b^2$, and we have neglected higher order terms of $O(\as^2)$ and $O(1/m_b^3)$. The parameter $\mug$ is renormalized at the scale $\mu=m_b$.  It is advisable to evaluate the QCD coupling constant at a scale lower than $m_b$. Here and in the following we adopt  \(\alpha_s = 0.25 \), which implies that the $O(\alpha_s )$ correction increases the \(\mu_G^2 \) coefficient by about  7\%.  Neglecting again higher order effects, the mean lepton energy is given by
\be
\langle E_\ell \rangle = 1.41 {\rm GeV} \left[\left( 1-0.02\, \frac{\as}{\pi}\right)\left(1 + \frac{\mu_\pi^2}{2m_b^2}\right) - \left(1.19 +4.20 \, \frac{\as}{\pi}\right)\frac{\mu_G^2(m_b)}{m_b^2} \right],\nonumber
\ee
while the variance of the lepton energy distribution is $\ell_2=\langle E_\ell^2 \rangle - \langle E_\ell \rangle^2$,
\be
  \ell_2= 0.183\, {\rm GeV^2} \left[ 1-0.16\, \frac{\as}{\pi} +\left(4.98\! -\!0.37\, \frac{\as}{\pi} \right)  \frac{\mu_\pi^2}{m_b^2} - \left(2.89 +8.44 \, \frac{\as}{\pi}\right)\frac{\mu_G^2(m_b)}{m_b^2} \right].\nonumber
\ee
In the  two above leptonic moments the NLO corrections to the coefficients of \( \mu_G^2 \)  are  larger than in the total rate: they  amount to \(+28 \%\) and \(+23 \%\), respectively. They have therefore  the same sign and size of the corrections to the width and photon energy moments in $b\to s \gamma$ \cite{ewerth}. 
Of course, the coefficients of the $O(\as)$ corrections   depend on the perturbative scheme and on the renormalization scale of $\mu_G^2$. In the kinetic scheme with cutoff $\mu_{kin}=1$GeV, which is often employed in semileptonic fits \cite{hfag,newfit}, the width becomes
%(see \cite{nnlo} and Refs. therein)
\be
\Gamma_{B\to X_c \ell \nu}=\Gamma_0 \left[ 1-1.11\, \frac{\as}{\pi} -  \left(\frac12- 0.99\, \frac{\as}{\pi}\right)\frac{\mu_\pi^2}{m_b^2} - \left(1.94 +3.46  \,\frac{\as}{\pi}\right)\frac{\mu_G^2(m_b)}{m_b^2} \right],\label{wikin}
\ee
where the NLO corrections to the coefficients of $\mupi,\mug$ are both close to 15\% but have different signs.\footnote{In the kinetic scheme the $O(1/m_b^3)$  corrections (here neglected) contribute to  the determination of  the perturbative corrections and  slightly modify the 
numerical values reported in Eqs.(\ref{wikin}-\ref{l2kin}).}
  Overall, the $O(\as\Lambda^2_{\rm QCD}/m_b^2)$ contributions 
decrease the total width by about 0.3\%. However, NLO corrections also modify the 
coefficients of  $\mupi,\mug$ in the moments which are fitted to extract the non-perturbative parameters, and will ultimately shift the values of $\mupi,\mug$ to be employed in (\ref{wikin}). Therefore, in order to quantify the eventual numerical impact of the new corrections on the semileptonic width and on $|V_{cb}|$, a new global fit has to be performed.
\begin{figure}[t]
\begin{center}
\includegraphics[width=9cm]{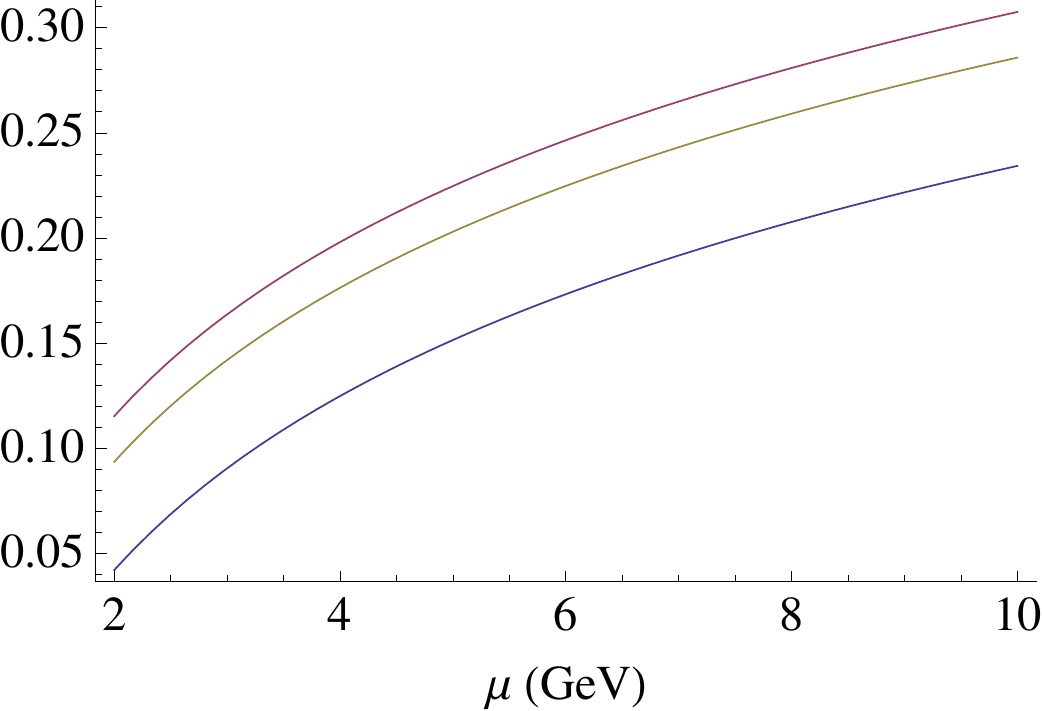}
\caption{\sf Relative NLO correction to the $\mug$ coefficients in the width (blue), first (red) and second central (yellow) leptonic moments as a function of the renormalization scale $\mu$ of $\mug$.}
\label{fig2}
\end{center}
\end{figure}

For what concerns the first  leptonic moment in the kinetic scheme we find
\be
\langle E_\ell \rangle = 1.41 {\rm GeV} \left[ 1-0.01 \,\frac{\as}{\pi} +\left(\frac12 -0.44\, \frac{\as}{\pi} \right)  \frac{\mu_\pi^2}{m_b^2} - \left(1.19 +3.21 \, \frac{\as}{\pi}\right)\frac{\mu_G^2(m_b)}{m_b^2} \right],\label{l1kin}
\ee
where the new corrections lead to a  $\approx0.5\%$ suppression. In practice, experiments measure this observable applying a lower cut on the lepton energy and the typical 
experimental error is lower than 0.5\%. We postpone the consideration of cuts to a future publication. In Eq.~(\ref{l1kin}) the $O(\as \Lambda_{\rm QCD}^2/m_b^2)$ correction is dominated by the term proportional to $\mug$, corresponding to a 20\% increase of the $\mug$ coefficient. Finally, the second central moment in the kinetic scheme is given by  
\be
\ell_2  = 0.183\, {\rm GeV^2} \bigg[ 1-0.24 \,\frac{\as}{\pi} +\left(4.98 -3.89\, \frac{\as}{\pi} \right)  \frac{\mu_\pi^2}{m_b^2} - \left(2.89 +7.01 \, \frac{\as}{\pi}\right)\frac{\mu_G^2(m_b)}{m_b^2} \bigg].\label{l2kin}
\ee
Here the new corrections lead to a 1.5\% suppression, again of the same order of the experimental error. The  
NLO correction to the $\mug$ coefficient is also about 20\%.

The size of the $O(\as \mug/m_b^2)$ corrections depends on the renormalization scale $\mu$ of the chromomagnetic operator. This is illustrated in Fig.\,\ref{fig2}, where the size of the NLO correction relative to the tree level results is shown for the width and the first two leptonic moments at different values of $\mu$. The NLO corrections are quite small for $\mu\approx 2$GeV and, as expected,  increase with $\mu$. For $\mu\gsim m_b$ the running of $\mug$ appears to  dominate  the NLO corrections.

\section{Summary}
We have calculated the $ O(\alpha_s) $ corrections to the Wilson coefficients of the chromomagnetic operator in inclusive semileptonic B decays, employing the techniques developed in Refs.\,\cite{ewerth} and \cite{Alberti:2012dn}.
This calculation turned out to be significantly  more demanding than that of \cite{ewerth},
 motivating us to explain the matching procedure in greater detail.
 We have also studied the numerical relevance of the new contributions in the absence of cuts:
 the perturbative $O(\as)$  corrections increase  the  \(\mu_G^2 \) coefficients in the total 
 semileptonic rate and in the first two leptonic moments by   \(15\%\) to \(20 \%\) if $\mug$ is 
 renormalized at $\mu=m_b$. For $\mu=2$GeV the corrections are in the 5-10\% range. The 
 complete $O(\as \Lambda_{\rm QCD}^2/m_b^2)$ correction to the width is a few
 per mill, but the corrections to the first two leptonic moments are of the same order of the experimental errors. A complete estimate of  the  effect of these corrections on the width and on $|V_{cb}|$ therefore requires their inclusion  in the global fit to the moments, which will be the subject of  a future publication.
 
\subsection*{Acknowledgements}
PG is grateful to G. Ridolfi for collaboration in the early stage of this project.  This work is supported in part by  MIUR under contract 2010YJ2NYW$\_$006  and by Compagnia di San Paolo under contract ORTO11TPXK.
\section*{Appendix}
We provide results for the contributions proportional to either \( C_F\) or \(C_A \)
\bea
W^{(G,1)}_i =  W^{(G,1)}_{i,u} +\frac{2}{3}C_F W^{(G,1)}_{i,F}  +\frac{2}{3}C_A W^{(G,1)}_{i,A}.
\eea
The term \( W^{(G,1)}_{i,u} \) contains a few recurring structures
\bea
 W^{(G,1)}_{i,u} &=&  
\bigg[ 2 C_F (1-E_0 I_{1,0})\, w_i^{(G,1)} +\frac{C_A}3 \: p_i^{(G,1)}]  \left[ \frac{1}{\hat{u}^2} \right]_+ +\frac{1}{2}C_A W^{(G,0)}_{i} \ln \frac{\mu}{m_b}\nonumber\\&&
+\frac{2d_i^{(G,1)}}{3\hat{y}^2} (2C_F -C_A )(1-E_0 ) \big[ 2(1-E_0 )\; I_{1,0} +\ln \rho \bigg] \delta (\uh ) 
\eea
where we have set \( d_1^{(G,1)}=1-E_0\), \( d_2^{(G,1)}=0 \), \( d_3^{(G,1)}=1 \), \( p_1^{(G,1)} =-\lambda_0  \), \( p_2^{(G,1)} =8(1-E_0 ) \) and \( p_3^{(G,1)} = -4E_0 \). 
The \(\mu\) dependence originates in the \(\MS\) renormalization of \(\mu_G^2 \). The remaining expressions are
\bea
W^{(G,1)}_{1,F}& =& \left[\left( \frac{8\rho}{\lambda_0} (1+2E_0 \!-\!3\rho)+5\lambda_0 +2(1\!-\!2E_0 +5\rho ) \right) I_{1,0}\right. \nn\\
  && \left.
\ \ \ \ \ -\frac{8}{\lambda_0}(2\rho+E_0 (1\!-\!3\rho )) \!-\!\frac{2E_0}{\rho} (1+5\rho ) \right] \left[ \frac{1}{\uh} \right]_+ +D^{(G,1)}_{1,F} \delta (\hat{u} ) 
\nn\\
  &&-  \left[   \frac{2}{\hat{y}}  \left( 20E_0^4 - \rho +E_0 \rho (2+\rho )\!-\!5E_0^3 (1+2\rho ) -E_0^2 (2+5\rho ) \right) I_{1,0}  \right. 
\\
&&\left.\ \ \ -(8\rho \!-\!5E_0 \!-\!23 E_0^2 ) \!+2S (5E_0^2 +E_0 \!-\!2\rho )\!+\!\frac{\lambda_0}{4\hat{y}} \ln \rho \right] \delta' (\uh )+ R^{(G,1)}_{1,F} \nn
\eea
\bea
W^{(G,1)}_{1,A} &=& \left[ \frac{1}{2}(1+8E_0 -3\rho) I_{1,0} - 1-E_0 \left( \frac{3}{2\rho}-\frac{5}{2} \right)  \right]\left[ \frac{1}{\uh} \right]_+  + D^{(G,1)}_{1,A} \delta (\hat{u} ) \nn\\ 
&&+ \left[ \frac{\lambda_0}{2} -\frac{1}{2}\left( \frac{\lambda_0}{2} -E_0 \right) \ln \rho  -E_0 (E_0 +2E_0^2+\rho ) I_{1,0} \right]  \delta' (\uh ) + R^{(G,1)}_{1,A}
\eea
\bea
W^{(G,1)}_{2,F} \!\! &=& \!\!  \left[ \frac{8}{\lambda_0} \left( \frac{E_0}{\rho} \!+\!4\!-\! 5E_0 \right)(1\!-\!2E_0 ) \!+\!8 \! \left( \! 1\!-\!\frac{1\!-\!13\rho +2E_0 (1+5 \rho )}{\lambda_0} \right) I_{1,0} \right] \left[ \frac{1}{\uh} \right]_+  \nn\\ &&+  \big[ 8E_0 (3-5E_0 ) I_{\Delta}\! -\! 2 (5E_0^2 -2E_0 \!-\!3\rho )I_{1,0} + (17-30E_0 )\ln\rho  +\!14 \!-\!26 E_0\big ]\delta' (\uh )  \nn\\&&
+ D^{(G,1)}_{2,F} \delta (\hat{u} ) \!+\! R^{(G,1)}_{2,F}
\eea
\bea
W^{(G,1)}_{2,A}\!\! &=&  \!\! \left[ \frac{4E_0}{\lambda_0 \rho}+\frac{4}{\lambda_0} (3+7\rho -11E_0 )  \! -\! \frac{3}{\rho}(1\! -\! 3\rho )  \! -\! 4  \left( \frac{1\! -\! 11\rho +E_0 (3+7\rho)}{\lambda_0}\! -\! 2 \right) I_{1,0} \right]\!\! \left[ \frac{1}{\uh} \right]_+
\nn\\ &&
\!\!\!\!\!\!\!\!\! +\left[ 2 (1\!-\!2E_0 )(1+E_0 )I_{1,0}   \!- \!4(1\!-\!E_0 )  +(1\!-\!2E_0 )\ln\rho  \right] \delta' (\uh )\!+\! D^{(G,1)}_{2,A} \delta (\hat{u} ) \!+\! R^{(G,1)}_{2,A}
\eea
\bea
W^{(G,1)}_{3,F}\!\! &=& \!\!  \left[2 \left( \frac{4}{\lambda_0} (E_0 (1-5\rho ) +3\rho )+5E_0 +2 \right) I_{1,0} -\frac{2}{\rho} - \frac{8}{\lambda_0} (1+3E_0 -5\rho) \right] \left[ \frac{1}{\uh} \right]_+ 
\\ &&
+ \big[20 E_0^2 I_{1,0} - 10 E_0 S -\frac{5}{2} \ln \rho   -4E_0 I_{\Delta}-3-25 E_0 \big] \delta' (\uh )+  D^{(G,1)}_{3,F} \delta (\hat{u} ) + R^{(G,1)}_{3,F}\nn
\eea
\bea
W^{(G,1)}_{3,A}\!\! &=& \!\!  \left[ 2 \left( \frac{E_0 +\rho (4-3E_0 )}{\lambda_0} +2 \right) I_{1,0} 	\!-\! \frac{2}{\lambda_0} (1+4E_0 \!-\! 3\rho ) \!-\! \frac{3\!-\! 7\rho }{2\rho}  \right]   \left[ \frac{1}{\uh} \right]_+ + R^{(G,1)}_{3,A}  \nonumber
\\ &&
+  \left[  E_0 - (E_0 +2E_0^2 +\rho )\; I_{1,0}-\left( E_0 -\frac{1}{2} \right) \ln \rho \right]\delta' (\uh )+ D^{(G,1)}_{3,A} \delta (\hat{u} ) 
\eea
We have called \(D_{i,F/A}^{(G,1)} \) the various coefficients of the \(\delta(\uh ) \) distribution
\bea
D^{(G,1)}_{1,F}&=&\left[ \!1+4E_0+ 5E_0^2 (1\!-\! 4E_0 )  \!-\!(9\!-\!8E_0 )\rho  +\frac{2}{\hat{y}}(1\!-\! E_0 ) (5E_0^2 \!-\! 4E_0 \!-\! 2)\right.
\nn\\&&
\ \ +\left.\frac{12 E_0^2}{\lambda_0} (1\!-\!E_0 )(1+3E_0 )\! \right] \! I_{1,0} 
+\frac{2}{\hat{y}}(1\!-\!E_0)\!-\!\frac{E_0}{2\rho} (1\!-\!20E_0 )\!-\!\frac{1}{2}(8\!-\!27E_0 \!-\!40E_0^2 ) 
\nn\\&&
  + \left[ \frac{2E_0}{\rho} \!-\!\frac{1}{2}(4\!-\!31 E_0)\!-\!(1\!-\!E_0 )\frac{1+5E_0 }{\hat{y}}   +\frac{4E_0}{\lambda_0} (2\!-\!E_0 )(1+3E_0 ) \right] \ln \rho 
\\&& 
-8\rho 
+\! \frac{4E_0}{\lambda_0}(1\!-\!E_0)(1+3E_0) \! 
+\left( 2\!-\!4E_0 \! +\! 5\lambda_0 \! +\! \frac{2\rho}{\lambda_0} (4\! +\! 8E_0 \! +\!5\lambda_0) \!-\!24\frac{\rho^2}{\lambda_0} \right)\! I_{\Delta} \;\;\;\;\nn
\eea
\bea
D^{(G,1)}_{1,A} &=&  \left[ \frac{1\!-\! E_0}{\hat{y}}(4\!-\! 5E_0 )\!-\! \frac{1}{2}(3\!-\! 18E_0 \!+\! 8E_0^2 \!-\! 3\rho)\! -\! \frac{2E_0^2}{\lambda_0}(1\!-\! E_0 )(1+3E_0 ) \right]\! I_{1,0} \nonumber
\\&&+
 \left( \frac{1}{4}(4-5E_0 )+\frac{3E_0}{2\rho}+\frac{3-5E_0}{2\hat{y}} \right) \ln \rho  +\frac{1}{2}(1+8E_0 -3\rho) I_{\Delta} \;\;\;\, 
 \nn\\&&
  +\frac{E_0}{2}(5+4E_0 )  -\frac{E_0}{2\rho}-2\rho-\frac{1-E_0}{\hat{y}}+\frac{2E_0}{\lambda_0}(1-E_0 )(1+3E_0 )\;\;\;\,
\eea
\bea
D^{(G,1)}_{2,F} &=&  
2  \left( 21E_0 -9 -20 E_0^2 -\frac{12}{\lambda_0} (1-E_0 )(1-E_0 -3E_0^2 ) \right) I_{1,0} \nn\\&&
+\frac{8}{E_0 \lambda_0}(1\!-\!E_0 )(1\!-\!9E_0 +11E_0^2 )\! -\!  \frac{4(1\!-\! 2E_0 )}{\lambda_0 \rho} (9\rho +E_0 (2\!-\! 5\rho )) \ln \rho  \nn\\  &&+\frac{2\!-\!17 E_0 +20 E_0^2}{E_0 \rho}\! -\!5(3\!-\!8E_0 )-  \frac{8}{\lambda_0} (1\!-\! \lambda_0 \!-\! 13\rho +2E_0 (1+5\rho )) I_{\Delta}
\eea
\bea
D^{(G,1)}_{2,A} &=& \left( 10E_0 -3 -\frac{4}{\lambda_0}(1-E_0 )(2+12E_0 -19E_0^2 ) \right) I_{1,0}-3+4E_0 \nn\\ &&-\frac{1}{\rho} - \left( 1+\frac{1-3E_0}{E_0 \rho}+ \frac{2}{E_0 \lambda_0}(2-E_0 )(1+4E_0 -7E_0^2 )\right) \ln \rho \nn\\ &&+\frac{4}{\lambda_0}(1-E_0 )(4-5E_0 )\! -\! \frac{4}{\lambda_0}(1-2\lambda_0 -11\rho +E_0 (3+7\rho )) I_{\Delta} 
\eea
\bea
D^{(G,1)}_{3,F} \!\!&=&\!\! \frac{8}{\lambda_0}(1+4E_0 \!-\! 4E_0^2)+ \left( \frac{35}{4}+\frac{2}{\rho}\!-\! \frac{6\! -\! 9 E_0 +5E_0^2}{2\hat{y} (1-E_0 )}+4(2\!-\! E_0 )\frac{1+3E_0 \!-\! 5E_0^2 }{\lambda_0 (1-E_0 )} \right) \ln \rho \nonumber\\ &&+\frac{2}{\hat{y}} +2  -(1-20 E_0 )\left( 1+\frac{1}{2\rho} \right)  +\frac{2}{\lambda_0}  (E_0 (4+5\lambda_0 -20\rho) +2(\lambda_0 +6\rho )) I_{\Delta} \nn\\ &&+
\left( 2(1+4E_0 -10 E_0^2 )-\frac{1}{\hat{y}}(8-9E_0 +5E_0^2 )+\frac{8E_0}{\lambda_0}(1+2E_0 -6E_0^2 ) \right) I_{1,0}  \;\;\;\;\;\;\;\;\;\;
\eea
\bea
D^{(G,1)}_{3,A} &=& \frac{2}{\lambda_0} (E_0 (1+8E_0 ) -\rho(4+3E_0 )) I_{\Delta}+2E_0 -\frac{1}{2\rho}-\frac{1}{\hat{y}}-\frac{4E_0^2}{\lambda_0} \nonumber\\ &&-\left( 1-\frac{3}{2\rho}+\frac{2-3E_0}{2\hat{y} (1-E_0)}-(2-E_0 )\frac{1+4E_0 -3E_0^2}{\lambda_0 (1-E_0)} \right) \ln \rho \nonumber\\ &&+\left( \frac{3}{2}(3-E_0)-\frac{1-3E_0}{\hat{y}}+\frac{4E_0}{\lambda_0}(1+4E_0 -2E_0^2 )\right) I_{1,0}
\eea
The terms labelled as \(R^{(G,1)}_{i,F/A}\) stand for the regular contributions 
\bea
R^{(G,1)}_{1,F} &=& \left[ \frac{4}{\lambda}(1\! -\! 3E+\rho)\! - \! \frac{2\!-15E+5\uh}{2} -\frac{24E_0 -15\lambda_0 -52\rho}{2\uh}+\frac{\uh}{\lambda} (11-13E) +\frac{5\uh^2}{\lambda} \right] I_1 \nn\\ &&
+ \frac{2E_0}{\hat{u} \rho}(1+5E_0 \!-\!5\rho)\! -\! \frac{\rho}{4z^3} (5\lambda +7z ) +\frac{12-11 E -13\rho +10\rho E}{\lambda} +\frac{13}{4} \left( 1+\frac{1}{z} \right) \nonumber\\ &&
-\frac{5}{2\rho z} (\lambda +2E \rho +\rho^2 ) - \left[ \frac{1}{\uh}\left( 2(1\! -\! 2E_0) +5(\lambda_0 +2\rho) \right) +\frac{8\rho}{\lambda_0 \uh} (1+2E_0 -3\rho)  \right]\! I_{1,0} \nn\\ &&+\frac{8E_0}{\lambda_0 \hat{u}} (1+2E_0 -3\rho) -\frac{5}{2\rho} \left( z+4(1-E) \right) +\frac{5}{8 z^2} (4E+\lambda -4E \rho -2 \rho^2)\;\;\;  \nonumber\\ &&-\frac{E}{\lambda z} (4\!-\!7\rho +5\rho^2)\!-\!\frac{z}{\lambda}(5E \!-\! 13)  +\frac{\lambda_0 (1+5E_0 )+4\rho (1+3E_0 )}{\uh^2} (I_1 -I_{1,0}) \;\;\;
\eea
\bea
R^{(G,1)}_{1,A} &=&   \frac{E}{2z^2}-\frac{3E\rho}{\lambda z}-\frac{6z}{\lambda}+ \left[ 1+\frac{1-2E}{\lambda}(8E +3\rho) +\frac{3z}{\lambda} (1+2E) \right] I_1  \nn\\ &&-\frac{3-\rho}{4\rho} +\frac{3E-\rho}{2\rho z}-\frac{8-13E-6\rho}{\lambda} +\frac{1+8E_0 -3\rho}{2\uh}(I_1 -I_{1,0})
\eea
\bea
R^{(G,1)}_{2,F} &=& - \left[ 25+\frac{48}{\lambda^2} (1-5E +8\rho-5E\rho+ \rho^2) +\frac{166-152E +74\rho}{\lambda}  \right. 
\nn\\ &&
+\frac{8}{\lambda \uh} (1-4E +3\rho) +\frac{4}{\uh} (6-5E)+\frac{10\uh}{\lambda}(19\! -\! 5E +\uh)+\frac{60\uh^3}{\lambda^2}
\nn\\ &&
\left.   -\frac{12\uh}{\lambda^2} (\!-\!39+47E \! -\!41\rho +13E \rho) +\frac{12\uh^2}{\lambda^2} (42\! -\! 23E +5\rho)  \right] I_1 
\nn\\ &&
-\frac{45}{2z}-\frac{4}{\rho \uh} (3-5E +5\rho) +\frac{\rho}{z^3} (8\!-\!10E \!-\!5\rho ) +\frac{2+10E-15\rho}{2z^2} 
\nn\\ &&
+\frac{12}{\lambda^2}\left( E (39\!-\!\rho )\!-\!20(1+\rho) \right)-\frac{8}{\lambda_0 \uh} (2-13E_0 +10\rho) 
\nonumber\\  &&
\!-\! \frac{12\uh^2}{\lambda^2} (23\!-\! 5E)+\frac{4}{\rho z} (4\!-\!5E) -\frac{2E}{\lambda z} (4-7\rho +5\rho^2 )\left( \frac{1}{z} -\frac{6}{\lambda} \right)    
\nn\\ &&
+\frac{106 E \!-\!199+10\rho \!-\!73\uh}{\lambda}  \!-\!\frac{8}{\lambda \uh}(4\!-\!3E) \!-\!\frac{12\uh}{\lambda^2} (47\!-\! 42E +13\rho ) 
\nn\\ &&
-\frac{4(5-16E)-3\rho (9-10E) +10\rho^2}{\lambda z} - \frac{8E_0}{\lambda_0 \rho \uh} -\frac{8E}{\lambda \rho} \left( \frac{1}{z}-\frac{1}{\uh} \right)
\\ &&
  +\frac{8}{\lambda_0 \uh} (1+2E_0 -\lambda_0 -13\rho +10 E_0 \rho )I_{1,0} -\frac{8E_0}{\uh^2} (3-5E_0) (I_1 \! -\! I_{1,0})\nn
\eea
\bea
R^{(G,1)}_{2,A} \!\! &=& \!\!  \left[ \frac{2}{\lambda} (61\! -\! 52E +25\rho +40\uh)  - \frac{24}{\lambda^2} \left( 2E(2+7\rho )-\rho (13+5\rho ) \right)+\frac{12\uh^2}{\lambda^2} (25\!-\!6E)   \right.
 \nn\\ &&
\;\;\;\; \left. +\frac{8}{\uh}-\frac{4}{\lambda \uh} (1+3E-11\rho +7E\rho ) +\frac{12\uh}{\lambda^2} (29\!-\!40E +35\rho\! -\!6E \rho )  \right] I_1 
 \nonumber\\  &&
 +\frac{4}{\uh \lambda_0}  \left( 1+E_0 (3-8E_0 ) -\rho (3-7E_0 ) \right) I_{1,0}  +\frac{6\uh}{\lambda} \left( 1+\frac{12\uh}{\lambda} \right)
\nn\\ && 
  +\frac{4}{\lambda z}(2\!-\!5E\!-\!3\rho) +\frac{4}{\lambda \uh} (3\!-\!11E +7\rho) +\frac{12\uh}{\lambda^2} (40-25E +6\rho)
 \nn\\ &&
 +\frac{1}{z^2} -\frac{4E}{\lambda \rho z}   +\frac{12}{\lambda^2} \left( 8-29E +2\rho (14-5E) \right)- \frac{4}{\rho \uh} \left( \frac{E_0}{\lambda_0} -\frac{E}{\lambda} \right)
\nn\\ && 
  -\frac{4}{\lambda_0 \uh} (3+7\rho -11E_0)  +\frac{6}{\lambda} (24-5E) -\frac{6E\rho}{\lambda z} \left( \frac{1}{z}-\frac{6}{\lambda} \right)+\frac{3}{\rho z}
\eea
\bea
R^{(G,1)}_{3,F} \!\!&=&\!\!  - \left[ \frac{25}{2} +\frac{6}{\lambda} (2-4E +3\rho ) +\frac{8\rho}{\lambda \uh} +\frac{4}{\uh} (1-5E) +\frac{2\uh}{\lambda} (14\! -\! 5E) \! -\! \frac{4E_0}{\uh^2} (1+5E_0) \right] I_1 
\nn\\ &&
 - \left[ \frac{4E_0 +5\lambda_0 +20\rho}{\uh^2} +\frac{8}{\lambda_0 \uh} (E_0 +3\rho -5E_0 \rho) +\frac{2}{\uh}(2+5E_0 ) \right] I_{1,0}
\nn\\ &&
+\frac{8}{\lambda_0 \uh} (1+3E_0 -5\rho) -\frac{4}{\lambda} (6-7E)  +\frac{2}{\rho \uh} (5E-5\rho +1) -\frac{10\uh}{\lambda} 
\nn\\ &&
+\frac{8E}{\lambda \uh}-\frac{5E\rho}{z^3} +\frac{5}{2z^2} (1+E-\rho )   -\frac{5}{z}  +\frac{2E}{\lambda z} (2-5\rho) -\frac{10E}{\rho z}
\eea
\bea
R^{(G,1)}_{3,A} &=&  \left[ \frac{2}{\lambda} \left( 3z +5(1-2E ) \right) +\frac{2}{\lambda \uh} (E+4\rho -3E\rho ) +\frac{4}{\uh} \right] I_1    \nn\\ &&  + \frac{2}{\lambda_0 \uh} (1+4E_0 -3\rho)  \!-\! \frac{2}{\lambda_0 \uh}  (E_0 +2\lambda_0 +4\rho -3E_0 \rho ) I_{1,0} \nonumber\\&&
+\frac{3}{2\rho z}\! -\! \frac{2}{\lambda \uh}(1+4E -3\rho )+\frac{2}{\lambda}(10-3E) +\frac{1}{2z^2} -\frac{2E}{\lambda z}
\eea
where we have introduced $z=\hat u + \rho$ and  $\lambda=4(E^2-\rho-\hat u)$. The integrals 
$I_1$, $I_{1,0}$, $I_{2,0}$, and $I_{4,0}$ are given in the Appendix of \cite{Alberti:2012dn},
and $I_{\Delta} =I_{2,0}-I_{4,0} $. The plus distributions are defined by their action on a test function $f(\hat u)$:
\be
\label{plusn}
\int d\uh \left[\frac{1}{\uh^m}\right]_+  f(u)=\int^{1}_0 d\uh\,
\frac{1}{u^m} \left[f(u)-\sum_{p=0}^{m-1} \frac{u^p}{p!} f^{(p)} (0)\right]
\ee
with $f^{(p)}(u)= \frac{d^p f(u)}{d u^p}$.

\end{document}